\documentclass[a4paper]{article}
\usepackage[utf8]{inputenc}
\usepackage[english]{babel} 
\usepackage{amsmath,amssymb,amsthm,mathbbol} 
\usepackage[margin=2cm]{geometry}
\usepackage{esint}
\usepackage[colorlinks]{hyperref}
\usepackage{enumerate}
\usepackage{framed,comment}
\usepackage{upgreek}
\usepackage{pgfplots}
\usepackage{float}
\usepackage{tikz,tikz-cd}
\usepackage{rotating}
\usepackage[color=yellow]{todonotes}
\usepackage{cancel}
\usepackage{caption}
\usepackage{subcaption}
\usepackage{circuitikz}
\usepackage{multirow}
\usepackage{authblk}

\RequirePackage[authoryear]{natbib}

\newtheorem{theorem}{Theorem}[section]

\newtheorem{remark}[theorem]{Remark}

\usepackage{parskip}
\numberwithin{equation}{section}

\newcommand{\el}[1]{\textcolor{blue!100}{EL: #1}}

\makeatletter
\DeclareFontFamily{OMX}{MnSymbolE}{}
\DeclareSymbolFont{MnLargeSymbols}{OMX}{MnSymbolE}{m}{n}
\SetSymbolFont{MnLargeSymbols}{bold}{OMX}{MnSymbolE}{b}{n}
\DeclareFontShape{OMX}{MnSymbolE}{m}{n}{
    <-6>  MnSymbolE5
   <6-7>  MnSymbolE6
   <7-8>  MnSymbolE7
   <8-9>  MnSymbolE8
   <9-10> MnSymbolE9
  <10-12> MnSymbolE10
  <12->   MnSymbolE12
}{}
\DeclareFontShape{OMX}{MnSymbolE}{b}{n}{
    <-6>  MnSymbolE-Bold5
   <6-7>  MnSymbolE-Bold6
   <7-8>  MnSymbolE-Bold7
   <8-9>  MnSymbolE-Bold8
   <9-10> MnSymbolE-Bold9
  <10-12> MnSymbolE-Bold10
  <12->   MnSymbolE-Bold12
}{}

\let\llangle\@undefined
\let\rrangle\@undefined
\DeclareMathDelimiter{\llangle}{\mathopen}%
                     {MnLargeSymbols}{'164}{MnLargeSymbols}{'164}
\DeclareMathDelimiter{\rrangle}{\mathclose}%
                     {MnLargeSymbols}{'171}{MnLargeSymbols}{'171}
\makeatother

\pgfplotsset{compat=1.14}

\usepackage{ulem}

\usepackage{color}

\newcommand{\rep}[2]{\textcolor{blue}{\sout{#1} #2}}
\newcommand{\rem}[1]{\textcolor{blue}{\sout{#1}}}
\newcommand{\add}[1]{\textcolor{blue}{#1}}
\newcommand{\comm}[1]{\textcolor{blue}{BJG: \textit{#1}}}

\title{A flexible numerical tool for large dynamic DC networks}
\author[1]{E. Luesink}
\author[2]{J.S. Giraldo}
\author[1,3]{B.J. Geurts}
\author[1]{J. Hurink}
\author[1,4]{H.J. Zwart}
\affil[1]{Department of Applied Mathematics, Faculty EEMCS, University of Twente, PO Box 217, 7500 AE Enschede, The Netherlands}
\affil[2]{Energy Transition Studies Group, Netherlands Organisation for Applied Scientific Research (TNO), Netherlands}
\affil[3]{Center for Computational Energy Research, PO Box 6336, 5600 HH Eindhoven, The Netherlands}
\affil[4]{Department of Mechanical Engineering, Eindhoven University of Technology, PO Box 513, 5600 MB Eindhoven, The Netherlands}

\date{\today}

\begin{document}

\maketitle

\begin{abstract}
DC networks play an important role within the ongoing energy transition. In this context, simulations of designed and existing networks and their corresponding assets are a core tool to get insights and form a support to decision-making. Hereby, these simulations of DC networks are executed in the time domain. Due to the involved high frequencies and the used controllers, the equations that model these DC networks are stiff and highly oscillatory differential equations. By exploiting sparsity, we show that conventional adaptive time stepping schemes can be used efficiently for the time domain simulation of very large DC networks and that this scales linearly in the computational cost as the size of the networks increase. 
\end{abstract}

\section{Introduction}

\subsection{Motivation}
In 2019, the European Green Deal introduced targets to address climate change in the European Union. The climate target plan lists a 55\% reduction of greenhouse gas emissions in 2030 compared to the 1990 levels. Hereby, the production and use of energy across economic sectors account for more than 75\% of the EU's greenhouse gas emissions. Therefore, in order to achieve emission reduction, the power sector must be transformed into a system that is largely based on renewable sources, complemented by the rapid phasing out of decarbonising gas and coal, while still being secure, reliable and affordable. However, renewable sources introduce a high amount of uncertainty into the existing electricity grid and also shift the focus from solely AC networks to hybrid AC-DC networks. 

In the context of such AC-DC networks, inverters play an important role. In \cite{park2020inverter} it is shown that that DC-AC inverters can not achieve 100\% efficiency, which motivates using as few such devices as possible. Furthermore, \cite{garces2023port} discuss the control of such inverters. An implication of these researches is that in the future large networks of DC devices will likely occur that are connected via few inverters to AC networks. For AC networks there exist already several powerful simulation and optimisation tools in the frequency domain that assist analysis and design, such as \cite{eminoglu2010dspfap, krishnamurthy2016psst, brown2017pypsa, thurner2018pandapower}. However, these tools generally cannot be used directly for DC networks, as DC devices are modelled in the time domain and require the solving of differential equations. To aid the design of stable and reliable DC networks, novel simulation tools are required that themselves are fast and reliable. 

In this paper, we propose and analyse a flexible numerical tool for the simulation of DC networks of varying size. In particular, we can simulate networks consisting of thousands of nodes and edges for several seconds. The analysis includes a study of the performance of several adaptive time-stepping methods for the differential equations that model DC networks consisting of distributed generation units (DGUs) (\cite{trip2018distributed}). The considered network topologies are based on the IEEE power network and PEGASE test cases (\cite{josz2016ac}) and can be accessed with the Python package pandapower (\cite{thurner2018pandapower}). Hereby, the smallest networks consist of a couple of nodes and edges and the largest networks consist of thousands of nodes and edges. We show that the proposed method scales linearly in complexity when time integration is performed using explicit methods. The complexity scales nonlinearly when implicit methods are used and we show that this can be favourable. We show that the tool can be employed for communication optimisation between the controllers as well as fault simulation.

\subsection{Literature review}
DC networks and DC microgrids are a central theme in the energy transition since many modern sources and loads (e.g., photovoltaic cells, batteries, electronic appliances) can be directly connected to DC networks by using DC-DC converters. This enables DC microgrids to be more efficient than AC microgrids as noted by \cite{justo2013ac}. Hereby, the implementation of DC grids and networks is not limited to low voltage operation, but also has its importance and can be of use at the distribution level (for details see \cite{dragivcevic2015dc}). At the transmission level, a large amount of research is devoted to the study and development of high-voltage DC (HVDC), because of its beneficial long-distance bulk-power delivery, asynchronous interconnections and long submarine cable crossings, (\cite{bahrman2007abcs}). The long-distance bulk-power capabilities of HVDC gets especially important in the interconnection of large offshore wind farms with mainland transmission systems. 

Due to this increasing relevance of DC grids, simulation tools for these networks are important assets to support the ongoing transition. The seminal work \cite{milano2013python} proposes a simulation tool for power system analysis in the time domain. In the present work we propose a simulation tool (also implemented in Python) that can deal with the wide range of involved frequencies present in DC networks, and which has a good performance also for large networks.

The stable operation of DC networks requires dedicated control algorithms of which many were developed in the last decade, see \cite{nasirian2014distributed, zhao2015distributed, han2017distributed, prabhakaran2017novel, tucci2018stable, de2018power, trip2018distributed, trip2018distributed2, cucuzzella2018robust, garces2019stability, kosaraju2020differentiation}. In most of these works, the main objectives of the developed control algorithms are two-fold. The first objective is proportional current sharing, where the total current that has to be generated is allocated according to local generation capabilities and local demand. The second objective is average voltage regulation, which aims to operate the network on average at a desired voltage. In the development of these control schemes, particular attention is paid to distributed control, enabling the control schemes to be scalable, have Plug-and-Play capabilities and are able to react quickly to changes in load. The proposed control schemes, on the one hand, can be fully local, meaning that the controller reacts solely on local measurements, but can also be centralised, where measurements are shared via a communication network. A further useful property of the proposed control schemes together with the dynamical model of a DC network is that of passivity, introduced in \cite{willems1976realization}. This passivity property guarantees that the interconnection of two passive systems is again passive, which means that the stable operation of a DC network becomes independent of its topology. 

\subsection{Contributions}
The main contribution of this paper is a flexible numerical tool that
\begin{itemize}
    \item solves differential equations modelling large DC networks in the time domain,
    \item has a computational cost that scales linearly in the spatial complexity,
    \item and can be used for the design and optimisation of DC networks.
\end{itemize}

In Section \ref{sec:modelling}, we briefly discuss the graph theoretic treatment of the topology of the networks. The objective of this section is to illustrate how the incidence matrix can be used as a bridge between linear algebra, graphs and topology. It allows for an elegant representation of the Kirchhoff laws and Tellegen's theorem. Since electrical networks are typically sparse, the incidence matrix is also sparse. Using this property is crucial to achieve computational efficiency when dealing with large networks.

In Section \ref{sec:dcmicrogrids}, we introduce the DC network model presented in \cite{trip2018distributed}. This model uses distributed generation units (DGUs) that determine the dynamics of the nodes of the network. The lines correspond to edges in the network and are modelled with the $\Pi$-model. Together with the controller proposed in \cite{trip2018distributed}, this allows to operate a DC network in a globally stable manner and thereby supporting both current sharing and average voltage regulation. These control objectives can also be achieved for large DC networks.

In Section \ref{sec:numerics}, we perform numerical simulations with five different time integration methods of the DC network model for the available IEEE and PEGASE benchmarks, see \cite{josz2016ac, thurner2018pandapower}. We show that the networks in most cases operate stable and conform to the control objectives of \cite{trip2018distributed} independent of their topology. In particular, we show that four out of the five methods show a linear increase in computational time as the dimension of the DC network increases. Finally, also the quality and computational complexity of the simulations are discussed. The paper ends with an outlook and conclusions in Section \ref{sec:outlookandconclusion}.

\section{Circuits and networks}\label{sec:modelling}
In this section, we discuss the graph theoretic concepts that are necessary for the modelling of electrical networks and circuits. Hereby, we carefully distinguish between a circuit and a network. Specific circuits are used to represent devices and a network is built out of devices. Thus for every device there is a corresponding circuit, but not every circuit corresponds to a device. The mathematical model for both circuits and networks is an undirected graph $G=(V,\mathcal{E})$, where $V=\{1,\hdots,n\}$ is the node set and $\mathcal{E}\subset V\times V$ is the set of undirected edges. Without loss of generality, we assume that for an edge $(i,j)\in \mathcal{E}$, we have $i>j$.

\begin{remark}
For the set of nodes in an electrical circuit we associate to each node two vector spaces, i.e., $C_0^{\boldsymbol{I}}$ (the space of node currents $\boldsymbol{I}$) and $C_0^{\boldsymbol{V}}$ (the space of node voltages $\boldsymbol{V}$). Similarly, to the set of oriented edges we associate the vector spaces $C_1^{\boldsymbol{f}}$ (edge current flows $\boldsymbol{f}$) and $C_1^{\boldsymbol{u}}$ (edge voltage drops $\boldsymbol{u}$). As vector spaces, $C_0^{\boldsymbol{I}}$ and $C_0^{\boldsymbol{V}}$ are dual to each other and isomorphic to $\mathbb{R}^n$, where $n$ is the number of nodes, and $C_1^{\boldsymbol{f}}$ and $C_1^{\boldsymbol{u}}$ are dual to each other and isomorphic to $\mathbb{R}^m$, where $m$ is the number of edges. In many works, such as \cite{bollobas1998modern}, $C_0^{\boldsymbol{I}}$ is identified with $C_0^{\boldsymbol{V}}$ and $C_1^{\boldsymbol{f}}$ is identified with $C_1^{\boldsymbol{u}}$. Although mathematically this is of no concern, from a physical point of view this is somewhat problematic since $C_0^{\boldsymbol{I}}$ contains the node currents and $C_0^{\boldsymbol{V}}$ contains the node voltages which are measured in different physical units. By retaining the distinctions between the node spaces $C_0^{\boldsymbol{V}}$ and $C_0^{\boldsymbol{I}}$ and the edge spaces $C_1^{\boldsymbol{u}}$ and $C_1^{\boldsymbol{f}}$, the modelling language in this paper is similar to port-Hamiltonian modelling, see \cite{maschke1993port,tonso2023port,garces2023port} and Brayton-Moser modelling, see \cite{brayton1964theorya,brayton1964theoryb,smale1972mathematical}.
\end{remark}

The topology of the graphs is given by the incidence matrix, which is a linear representation of the incidence operator $\mathbb{R}^n\mapsto\mathbb{R}^m$. For a given node, the incidence operator outputs which edges are incident to this node. The transpose of the incidence matrix is then a linear representation of the coincidence operator. For a given edge, the coincidence operator outputs the nodes that are connected by that edge. Formally, this means that the graph $G$ is represented by the oriented incidence matrix $B\in\mathbb{R}^{n\times m}$ defined by
\begin{equation}
    B_{ie} = \begin{cases}
        +1 \quad &\text{ if } e=(i,j) \text{ for some } j\in V,\\
        -1 \quad & \text{ if } e =(j,i) \text{ for some } j\in V,\\
        0 \quad &\text{ otherwise. }
    \end{cases}
\end{equation}
The incidence matrix $B:C^{\boldsymbol{I}}_1\to C^{\boldsymbol{I}}_0$ and its transpose $B^T:C_0^{\boldsymbol{V}}\to C^{\boldsymbol{V}}_1$ transform information on edges to information on vertices and vice versa. Note that the incidence matrix in the context of electrical circuits maps edge currents to node currents and the transpose of the incidence matrix maps node voltages to edge voltage drops. The incidence matrix has the property that all its columns sum to zero, i.e., $\mathbb{1}_n^TB = \mathbb{0}_m^T$, where $\mathbb{1}_n$ is the $n$-vector of ones and $\mathbb{0}_m$ is the $m$-vector of zeros. The incidence matrix is used for modelling electrical circuits and electrical networks. 

\paragraph{Circuits.} In the graph representation of electrical circuits, the edges are two-terminal electrical components, such as resistors, inductors and capacitors. These components form the elements of the edge set $\mathcal{E}_1$. Note that for circuits that contain components with more than two terminals, such as transistors and amplifiers, different modelling strategies are required. The node set $V$ consists of the the points of interconnection between two-terminal components. This leads to a graph $G=(V,\mathcal{E}_1)$ which represents the circuit. Let $B$ be the associated incidence matrix. Note that for circuits, it is often convenient to introduce a ground node $0$ and a separate edge set $\mathcal{E}_0 = \{(i,0) \,\vert\, i\in V\}$ that connects every node $i$ to the ground. To represent directed flows, we define for each edge $(i,j)$ an oriented current flow $f_{ij}\in\mathbb{R}$ and an oriented voltage drop $u_{ij}\in\mathbb{R}$. A circuit with a ground node is represented by a graph $G_{circuit}=(V\cup \{0\}, \mathcal{E}_0\cup\mathcal{E}_1)$.  The incidence matrix $B_{circuit}$ of the circuit then takes the form
\begin{equation}
    B_{circuit} = \left(\begin{matrix}
        -\mathbb{1}^T_n & \mathbb{0}^T_m\\
        \mathbb{1}_{n\times n} & B
    \end{matrix}\right),
\end{equation}
where $B$ is the incidence matrix of the graph $G=(V,\mathcal{E}_1)$. The ground node is necessary for normalisation, i.e., to define the gauge with respect to which one measures the potential difference. In general, one chooses the ground to have zero potential. Note that the ground node is used also in modelling external current injection and loads.

\paragraph{Networks.} As the ground node has been introduced at the circuit level, it is not necessary to introduce additional nodes in the graph representing an electrical network. Note that in an electrical network, the nodes are grounded devices and the edges are two-terminal circuits without a ground node. Hence, in the context of electrical networks, the role of the incidence matrix is the same as in circuits. It describes the topology, but instead of relating voltages and currents between nodes and edges, it relates the state of devices and the state of lines interconnecting the devices. 

If direction is not important in a network, e.g., in modelling communication between agents, the topology of such a network can be described by the Laplacian matrix $\mathcal{L}$ (see for instance \cite{bollobas1998modern}) which relates to the incidence matrix by $\mathcal{L} = BB^T\in \mathbb{R}^{n\times n}$. Often the Laplacian matrix is weighted (with the weights being positive real numbers that model for instance the strength of communication). In such a case one has the Laplacian matrix $\mathcal{L}_W = BWB^T$, with $W$ a diagonal matrix containing the strictly positive weights. The Laplacian matrix is sometimes also called the Kirchhoff matrix. It is a square matrix for which the eigenvalues and eigenvectors can be determined and is therefore a central object in spectral graph theory. 

Using the introduced topological notions, in the next section, we describe circuits and networks using algebraic graph theory.

\section{DC networks}\label{sec:dcmicrogrids}
In this section we show how to build DC networks consisting of particular DC devices called distributed generation units (DGUs) and physical lines, using the language of algebraic graph theory. We first explain the individual building blocks of the network and then how to combine them. For the stable operation of such a DC network, suitable control laws are required. We introduce control objectives and control laws based on \cite{trip2018distributed}. These control laws feature a communication network. This means three different graphs are required to represent the controlled DC network: the electrical circuits modelling the DGUs, the physical network that links the DGUs and the communication network that represents the controllers that are exchanging information with each other. Through the use of algebraic graph theory, the model of the DC network can be expressed in terms of an affine system of differential equations. Analysis of this affine system suggests suitable numerical methods to solve these equations, which are the topic of the subsequent section.

\subsection{Building blocks for a DC network}
In this subsection we introduce the circuit diagrams that represent the devices in the DC network. Let $G_{phys}$ denote the graph consisting of $n$ nodes and $m$ edges that represents the physical electrical network and its components. This graph will be used as the blueprint for assembling the DC network. 

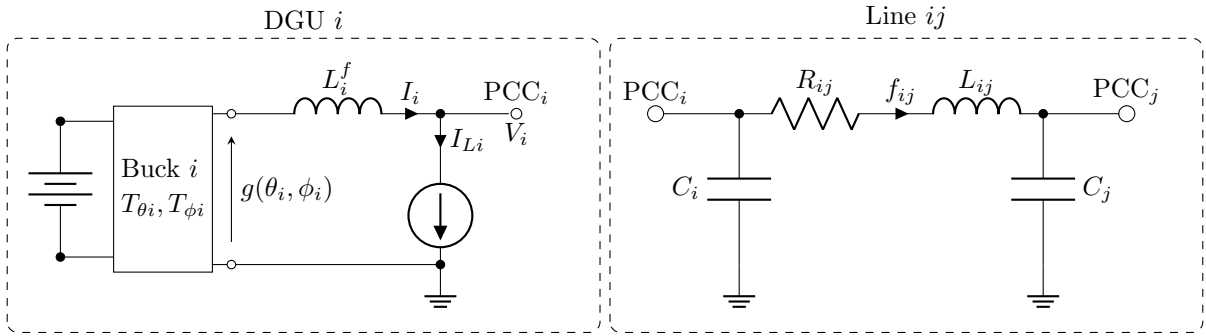
\begin{figure}[H]
\begin{center}
\begin{circuitikz}[american]
\draw (-2,0.1)
to[battery,invert,*-*] (-2,1.9);
\draw (-2,0.1)
to[short] (-1.3,0.1);
\draw (-2,1.9)
to[short] (-1.3,1.9);
\draw (-1.3,-0.1) rectangle node{\parbox{1.1cm}{Buck $i$\\$T_{\theta i}, T_{\phi i}$}} (0,2.1);
\draw (0,2) to[short,-o] (0.25,2);
\draw (0,0) to[short,-o] (0.25,0);
\draw [-stealth](0.25,0.3)--(0.25,1.7);
\draw (0.25,1) node[anchor=west]{$g(\theta_i,\phi_i)$};
\draw (0.3,2)
to[L=$L^f_{i}$,i=$I_{i}$,-*] (3,2)
to[short,i=$I_{Li}$] (3,1.3) 
to[isource,-*] (3,0) node[ground]{}
to[short] (0.3,0);
\draw (3,2)
to[short] (3.9,2);
\draw[rounded corners,dashed] (-2.7,-0.9) rectangle (5.1,3) {};
\draw (1.2,3) node[anchor=south]{DGU $i$};
\draw (4,2) circle (2pt) node[anchor=south]{PCC$_i$} node[anchor=north]{$V_i$};
\end{circuitikz}
    \begin{circuitikz}[american]
        \draw (-2.1,2) circle (3pt) node[anchor=south]{PCC$_i$};
        \draw (-2,2)
        to[short,-*] (-1,2) node[anchor=south]{};
        \draw (-1,2)
        to[C,l_=$C_i$] (-1,0) node[ground]{};
        \draw (-1,2)
        to[R=$R_{ij}$] (1,2)
        to[short,i=$f_{ij}$] (1.25,2)
        to[L=$L_{ij}$,-*] (3,2) node[anchor=south]{};
        \draw (3,2)
        to[C,l^=$C_j$] (3,0) node[ground]{};
        \draw(3,2)
        to[short] (4,2);
        \draw (4.1,2) circle (3pt) node[anchor=south]{PCC$_j$};
        \draw[rounded corners,dashed] (-2.7,-0.9) rectangle (5.1,3) {};
        \draw (1.2,3) node[anchor=south]{Line $ij$};
    \end{circuitikz}
\end{center}
\caption{The circuit on the left represents distributed generation unit (DGU) $i$ and the circuit on the right represents line $ij$. Line $ij$ connects to DGU $i$ by means of the point of common connection (PCC) indexed by $i$.}
\label{fig:blocks}
\end{figure}

Figure \ref{fig:blocks} shows the circuit diagram for the devices that serve as the building blocks of the DC network. Hereby, node $i$  ($i\in\{1,\hdots,n\}$) in $G_{phys}$ represents DGU $i$ and edge $(i,j)$ in $G_{phys}$ represents line $ij$ between DGUs $i$ and $j$. The topology of the network $G_{phys}$ can also be represented by an incidence matrix $B$.

The left circuit diagram in Figure \ref{fig:blocks} represents a DGU. A DGU is a controllable device and is used to model local current generation and local constant current loads. The state of DGU $i$ is given by the corresponding current $I_i$ and the voltage $V_i$. The control of a DGU is represented by a control law $g(\theta_i,\phi_i)$ and depends on control inputs $\phi_i$ (for local control) and $\theta_i$ (for nonlocal control) which determine the action of buck converter Buck $i$. Hereby, a buck converter is a device that can decrease the voltage while increasing the current depending on its control inputs. Hence, each DGU has four variables: the generated current $I_i$, the voltage $V_i$ and two control variables $\phi_i$ and $\theta_i$. Furthermore, DGU $i$ has several properties specified by parameters, which are the filter inductance $L^f_{i}$, the current demand $I_{L_i}$ and the strengths of the control inputs $T_{\phi i}$ and $T_{\theta i}$. More information on the control objectives and the control law $g(\theta_i,\phi_i)$ satisfying these objectives is given in the next subsection.

The right circuit diagram in Figure \ref{fig:blocks} represents a line and presents the well-known $\Pi$-model, which models the effects of charging currents, losses and electromagnetic waves present in underground cables, wires and transmission lines. In the remainder, we use the term lines to refer to these type of cables. The state of a line $ij$ is determined by the line current $f_{ij}$ and the voltage drop across the line $u_{ij}=V_i-V_j$, which can be determined from the voltages $V_i$ at DGU $i$ and $V_j$ at DGU $j$. The parameters of the line are the line resistance $R_{ij}$ and the line inductance $L_{ij}$. 

\begin{figure}[H]
\begin{center}
\begin{circuitikz}[american]
\draw (-2,0.1)
to[battery,invert,*-*] (-2,1.9);
\draw (-2,0.1)
to[short] (-1.3,0.1);
\draw (-2,1.9)
to[short] (-1.3,1.9);
\draw (-1.3,-0.1) rectangle node{\parbox{1.1cm}{Buck $i$\\$T_{\theta i}, T_{\phi i}$}} (0,2.1);
\draw (0,2) to[short,-o] (0.25,2);
\draw (0,0) to[short,-o] (0.25,0);
\draw [-stealth](0.25,0.3)--(0.25,1.7);
\draw (0.25,1) node[anchor=west]{$g(\theta_i,\phi_i)$};
\draw (0.3,2)
to[L=$L^f_{i}$,i=$I_{i}$,-*] (3,2)
to[short,i=$I_{Li}$] (3,1.3)
to[isource,-*] (3,0)
to[short] (0.3,0);
\draw (3,2)
to[short] (3.8,2);
\draw (3.9,2) circle (2pt);
\draw (3.9,2) circle (3pt) node[anchor=south]{PCC$_i$} node[anchor=north]{{\hspace{2em}}$V_i$};
\draw (4,2) to[short,-*] (5,2);
\draw (5,2)
to[C,l_=$C^L_{i}$,-*] (5,0) node[ground]{}
to[short] (3,0);
\draw (5,2)
to[short] (5.6,2);
\draw (5.6,2)
to[short] (6.5,2);
\draw (6.5,2)
to[R=$R_{ij}$] (8,2)
to[short,i=$f_{ij}$] (8.5,2)
to[L=$L_{ij}$] (10,2);
\draw (10,2)
to[short] (11,2);
\draw[rounded corners,dashed] (-2.7,-0.9) rectangle (6.1,3) {};
\draw (1.7,3) node[anchor=south]{DGU $i$};
\draw[rounded corners,dashed] (6.3,-0.9) rectangle (10.3,3) {};
\draw (8.3,3) node[anchor=south]{Line $ij$};
\end{circuitikz}
\end{center}
\caption{Circuit diagram describing DGU $i$ and the physical line connecting DGU $i$ to a DGU $j$. Note that the capacitor $C^L_i$ is the lumped capacitor that represents all parallel capacitors at the $i$th point of common connection.}
\label{fig:diagram}
\end{figure}
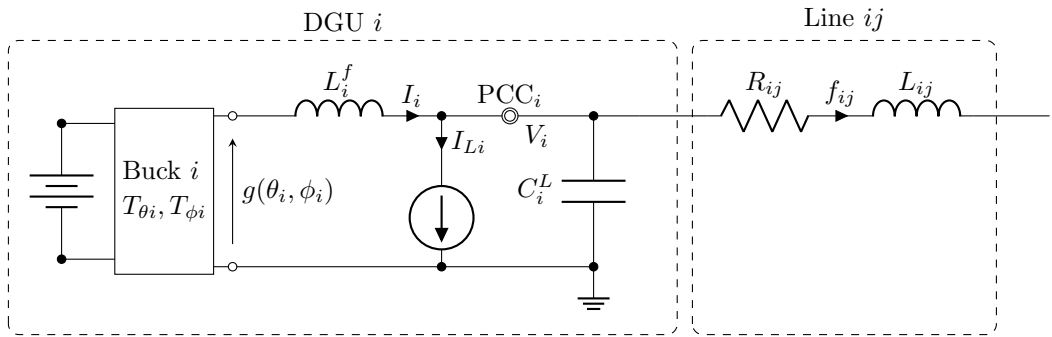

A network of DGUs interconnected by lines can model electricity supply and demand on a local and a global level. Line $ij$ is connected to DGU $i$ at the common point of connection PCC$_i$ of DGU $i$ and PCC$_j$ of DGU $j$. However, note that when DGU $i$ connects to several other DGUs there will be several parallel capacitors at the point of common connection. Since parallel capacitors can be replaced by a single equivalent capacitor $C^L_i$ with a capacitance equal to the sum of the capacitances of the parallel capacitors, it is convenient to make this equivalent capacitor $C^L_i$ part of the circuit representing the DGU. This leads to the circuit diagram of a DGU with a line as in \cite{trip2018distributed}, (see Figure \ref{fig:diagram}).

Using the building block given in Figure \ref{fig:diagram}, we can assemble the overall circuit diagram of a given DC network. In the next subsection, we consider desirable behaviour of DC networks and how to select the control inputs $\theta_i$ and $\phi_i$ to achieve this behaviour. 

\subsection{Control objectives}
In this subsection we consider desirable behaviour of DC networks. A DC network is a dynamical system whose state should conform to a given DC network policy. Such policies are usually very different between AC and DC networks. Generally, policies for AC networks tend to tightly restrict fluctuations in frequencies and harmonic distortion and are more flexible to current and voltage fluctuations, whereas for DC networks current and voltage fluctuations are tightly controlled.

To describe the control objectives, we introduce some further notation. Let $\boldsymbol{I} = (I_1, \hdots, I_n)$ be the $n$-vector of generated currents in the DGUs, let $\boldsymbol{V} = (V_1, \hdots, V_n)$ be the $n$-vector of voltages at the points of common connection of the DGUs and let $\boldsymbol{f} = (f_1, \hdots, f_m)$ be the $m$-vector of line currents. To ensure desirable operation of the DC network, we formulate two control objectives following \cite{trip2018distributed}. Let $\boldsymbol{I}_L = (I_{L1},\hdots,I_{Ln})$ be the $n$-vector of loads and let $\mathbb{1}$ be a vector of ones. In the DC network, it is required that, in a steady state $(\overline{\boldsymbol{I}}, \overline{\boldsymbol{V}}, \overline{\boldsymbol{f}})$, the total demand $\mathbb{1}^T \boldsymbol{I}_L$ of the network is shared among all the DGUs. Mathematically, this is formulated as $\mathbb{1}^T \overline{\boldsymbol{I}} = \mathbb{1}^T\boldsymbol{I}_L$. 

However, to avoid overusing a single source and to improve generation efficiency, DGUs need to share current proportional to their generation capacity. This is called the proportional current sharing objective. The generation capacity of a DGU is specified using weights and is formulated mathematically locally as $w_i \bar{I}_{i} = w_j\bar{I}_{j}$ for all $i,j\in V$, where $(w_1,\hdots,w_n)$ are the given strictly positive weights. A large weight $w_i$ corresponds to a relatively small generation capacity of DGU $i$. Upon combining the local formulation of the proportional current sharing objective above with the steady state requirement $\mathbb{1}^T\overline{\boldsymbol{I}} = \mathbb{1}^T\boldsymbol{I}_L$, we obtain the proportional current sharing objective in a global formulation.

\paragraph{Proportional current sharing.} Summarising, the proportional current sharing objective is then given by
\begin{equation}
    \|\boldsymbol{I}(t) - \overline{\boldsymbol{I}}\| \to 0 \quad \text{ with } \quad  \overline{\boldsymbol{I}} = W^{-1}\mathbb{1}\frac{\mathbb{1}^T \boldsymbol{I}_L}{\mathbb{1}^TW^{-1}\mathbb{1}},
\end{equation}
with $W={\rm diag}(w_1,\hdots,w_n), w_i>0$, for all $i\in V$.

For the second control objective considered, we assume that for DGU $i$ there exists a desired reference voltage $V_i^*$ and let $\boldsymbol{V}^*=(V_1^*,\hdots, V^*_n)$ be the $n$-vector of these desired reference voltages. In general, achieving proportional current sharing does not permit a steady state voltage. This means that it is in general not possible to achieve for the given desired voltages $\boldsymbol{V}^*$ a steady state voltage $\overline{\boldsymbol{V}}$. For this, average voltage regulation is chosen as an objective, where the weighted average $\mathbb{1}^TW^{-1}\overline{\boldsymbol{V}}$ of the voltages in the steady state $\overline{\boldsymbol{V}}$ is equal to the weighted average $\mathbb{1}^T W^{-1}\boldsymbol{V}^*$ of the desired reference voltages $\boldsymbol{V}^*$, with $W$ the same diagonal weight matrix as in the proportional current sharing objective. This follows the standard practice where the sources with largest generation capacity determine the grid voltage. In summary, the average voltage regulation objective is given by the following.

\paragraph{Average voltage regulation.} The average voltage regulation objective is formulated mathematically as
\begin{equation}
    \|\mathbb{1}^T W^{-1}\boldsymbol{V}(t)- \mathbb{1}^T W^{-1}\overline{\boldsymbol{V}}\| \to 0 \quad \text{ with } \quad \mathbb{1}^T W^{-1}\overline{\boldsymbol{V}} = \mathbb{1}^TW^{-1}\boldsymbol{V}^*.
\end{equation}

To achieve the proportional current sharing objective and the average voltage regulation objective, a suitable control law is required. However, both objectives are nonlocal since the proportional current sharing objective compares the steady state of DGU $i$ with the loads of other DGUs and the average voltage regulation objective uses the average voltage of all DGUs in the DC network. To account for this nonlocality, a communication network between DGUs is needed. For this let $G_{com}$ be the graph of the used communication network. The graph $G_{com}$ has the same node set as $G_{phys}$, but possibly a different edge set. Let $B_{com}$ denote the incidence matrix corresponding to $G_{com}$. \cite{trip2018distributed} show that the following control law $g(\theta_i,\phi_i)$ for DGU $i$ in a DC network achieves the control objectives
\begin{equation}\label{eq:controllaw}
    g(\theta_i,\phi_i) := -K_i(I_{i}-\phi_i)+w_i\sum_{j\in N_{com\,i}}\gamma_{ij}(\theta_i-\theta_j) + V_i^*,
\end{equation}
subject to
\begin{equation}\label{eq:controlde}
\begin{aligned}
        T_{\theta i}\frac{d}{dt}{\theta}_i &= -\sum_{j\in N_{com\,i}}\gamma_{ij}(w_i I_{i} - w_j I_{j}),\\
        T_{\phi i}\frac{d}{dt}{\phi}_i &= -\phi_i + I_{i}.
    \end{aligned}
\end{equation}

Hereby, $\theta_i$ and $\phi_i$ are the state variables for the controller, described by differential equations. Hereafter, we use the notation $\dot{\theta}_i$ to denote the time derivative of $\theta_i$. $T_{\theta_i}$ and $T_{\phi_i}$ are parameters that are used to tune the transient response of the controller. $N_{com\,i}$ is the set of neighbouring nodes of node $i$ in the communication network and $g(\theta_i,\phi_i)$ is the control input for buck $i$ of DGU $i$. The edge weight $\gamma_{ij}$ represents the ``strength" of communication across edge $ij$ of the communication network. This strength could for instance represent the quality of a certain connection, such as communication via WiFi or via UTP cables.  $K_i$ allows tuning of the transient response of the controller. 

By means of Kirchhoff laws, the constitutive relations for the components in the circuit in Figure \ref{fig:diagram} and the control law \eqref{eq:controllaw}, a system of differential equations can be formulated that describes the state of the DC network. In the next subsection, we formulate this system of differential equations by means of linear algebra and algebraic graph theory. 

\subsection{Linear representation}
In this section we introduce a compact formulation of the differential equations that model the DC network with control. To this end, we introduce the control input vectors $\boldsymbol{\theta}=(\theta_1,\hdots,\theta_n)$ and $\boldsymbol{\phi}=(\phi_1,\hdots,\phi_n)$ and for the current demand at each DGU the $n$-vector $\boldsymbol{I}_L = (I_{L1},\hdots, I_{Ln})$. The parameters of the DGU and the lines are assembled into diagonal matrices, $L^f = {\rm diag}_{i=1,\hdots,n}(L^f_{i})$ and $C^L = {\rm diag}_{k=1,\hdots,n}(C^L_{i})$ are the $n\times n$ diagonal matrices that contain, respectively, the inductances and capacitances of the DGUs. For the parameters of the buck converter, we set $T_\theta = {\rm diag}_{i=1,\hdots,n}(T_{\theta i})$, $T_\phi = {\rm diag}_{i=1,\hdots,n}(T_{\phi i})$ and $K={\rm diag}_{i=1,\hdots,n}(K_i)$. Furthermore, let $W={\rm diag}_{i=1,\hdots,n}(w_i)$ be the $n\times n$ diagonal matrix containing the weights that govern the relative generation capacity of the DGUs and for parameters of the lines, let $L = {\rm diag}_{k=1,\hdots,m}(L_{k})$ and $R = {\rm diag}_{k=1,\hdots,m}(R_k)$ be the $m\times m$ diagonal matrices that contain the line inductances and line resistances. Finally, by introducing the matrix $\Gamma={\rm diag}_{k=1,\hdots,m_{com}} \gamma_k$ governing the strength of communication and the weighted Laplacian matrix $\mathcal{L}_{com} = B_{com}\Gamma (B_{com})^T$ of the communication network we can represent the sums in \eqref{eq:controllaw}.

Having introduced the diagonal matrices containing the parameter values of the DGUs and the lines, and the matrices representing the various graphs, we formulate the system of differential equations that describes a DC network with control as follows
\begin{equation}
\begin{aligned}
L^f\boldsymbol{\dot{I}} &= -\boldsymbol{V}-K(\boldsymbol{I}-\boldsymbol{\phi}) + W\mathcal{L}_{com}\boldsymbol{\theta} + \boldsymbol{V}^*,\\
C^L\boldsymbol{\dot{V}} &= \boldsymbol{I} + B\boldsymbol{f} - \boldsymbol{I}_L,\\
L\boldsymbol{\dot{f}} &= -B^T \boldsymbol{V} - R\boldsymbol{f},\\
T_\theta\boldsymbol{\dot{\theta}} &= -\mathcal{L}_{com} W \boldsymbol{I},\\
T_\phi\boldsymbol{\dot{\phi}} &= -\boldsymbol{\phi} + \boldsymbol{I},
\end{aligned}
\label{eq:closedloop}
\end{equation}
The total number of differential equations for a DC network with control consisting of $n$ DGUs, $m$ lines and $m_{com}$ communication links is equal to $4n+m$. Note that the number of edges in the communication network has no influence on the number of differential equations. Instead, $m_{com}$ determines how many nonzero elements the Laplacian matrix $\mathcal{L}_{com}$ has. The system of differential equations \eqref{eq:closedloop} is affine. This means that \eqref{eq:closedloop} is of the form $\dot{\boldsymbol{x}} = A\boldsymbol{x} + \boldsymbol{b}$ with the state vector given by $\boldsymbol{x}=(\boldsymbol{I},\boldsymbol{V},\boldsymbol{f},\boldsymbol{\theta},\boldsymbol{\phi})^T$, the vector $\boldsymbol{b} = ((L^f)^{-1}\boldsymbol{V}^*,-(C^L)^{-1}\boldsymbol{I}_L,\mathbb{0}_m,\mathbb{0}_n,\mathbb{0}_n)$ and the matrix $A$ given by
\begin{equation}
A = \left(\begin{matrix}
    -(L^f)^{-1} K & -(L^f)^{-1} & \mathbb{0}_{n\times m} & (L^f)^{-1} W \mathcal{L}_{com} & (L^f)^{-1} K\\
    (C^{f})^{-1} & \mathbb{0}_{n\times n} & (C^{f})^{-1} B & \mathbb{0}_{n\times n} & \mathbb{0}_{n\times n} \\
    \mathbb{0}_{m\times n} & -L^{-1}B^T & -L^{-1} R & \mathbb{0}_{m\times n} & \mathbb{0}_{m\times n} \\
    -T_\theta^{-1}\mathcal{L}_{com} W & \mathbb{0}_{n\times n} & \mathbb{0}_{n\times m} & \mathbb{0}_{n\times n} & \mathbb{0}_{n\times n}\\
    T_\phi^{-1} & \mathbb{0}_{n\times n} & \mathbb{0}_{n\times m} & \mathbb{0}_{n\times n} & -T_\phi^{-1}
\end{matrix}\right)    
\label{eq:Jacobian}
\end{equation} 
The matrix $A$ is the Jacobian of the system and has dimension $(4n+m)^2$. By analysing $A$ one can deduce properties of the system of differential equations. We start by noting that $A$ has very few nonzero elements. The only nondiagonal submatrices in $A$ are the ones involving the matrices $B$ and $\mathcal{L}_{com}$ related to the topology of the physical network and the communication network. Hence, the sparsity of the networks determines to a great extent the sparsity of $A$. Further, we note that every row and every column of a Laplacian matrix sum to zero, hence $\mathcal{L}_{com}$ does not have full rank. It follows that $A$ is also not of full rank either and thus a singular matrix. This means that the system of equations \eqref{eq:closedloop} supports multiple equilibria, which is essential to satisfy the control objectives. Since the typical values of the inductance and capacitance for each DGU are small, the system has both very fast oscillations and slow evolution. 

In combination with the strong damping that the controllers introduce, the system can be stiff. Stiffness is a numerical phenomenon without a precise mathematical definition, but in practice it is characterised as follows. In the numerical solution of a set of differential equations, one would expect the step size to be small in regions where the solution curve varies a lot and the step size to be large in regions where the solution curve is close to having a flat slope. However, for some systems even in the case when the solution curve looks very smooth, the step size is still required to be unacceptably small. When this happens, the system is called stiff. 

To determine time efficient and accurate solution of stiff systems of differential equations specialised numerical methods are required. The slow-fast nature of the system indicates that adaptive time-stepping methods are appropriate, since for fixed-step methods the smallest possible time step has to be chosen for numerical stability. Given that the typical frequency of DC circuits is in the order of kilohertz to megahertz, solving \eqref{eq:closedloop} for one second would require thousands to millions of time steps. An alternative is to use adaptive methods, which can perform time stepping much more efficiently. However, the most widely used adaptive method for the numerical solution of differential equations and the workhorse for most mathematical software packages, the four-stage Runge-Kutta method, is not suitable for solving \eqref{eq:closedloop} due to stiffness, see \cite{butcher2000numerical, butcher2016numerical} and \cite{cash2003efficient}. In the next subsection, we discuss five standard adaptive numerical methods, including the four-stage Runge-Kutta method for performance comparison, as a means to solve the system of equations \eqref{eq:closedloop}.

\subsection{Adaptive time-stepping schemes}
In this section we describe the main advantages that adaptive solvers provide for DC networks, such as their ability to increase the time step size when the transient oscillations are subsiding. To solve \eqref{eq:closedloop}, we compare five adaptive time stepping methods: RK23, RK45, DOP853, BDF and Radau. These methods are the standard methods implemented in scipy's solver for initial value problems, \cite{virtanen2020scipy}. A detailed explanation of how such methods work and how to implement these methods can be found in \cite{hairer1993solving}. Before highlighting the specific properties of the methods, we first discuss the methods in a more general manner.

In theory, adaptive stepping methods can achieve arbitrarily large time steps when the solution is in equilibrium. The algorithms in adaptive time-stepping schemes usually allow gradual increases in the time step to retain time accuracy. This means that it takes a while to increase or decrease the time step size. For DC network simulations, particular care has to be taken to achieve time accurate solutions also in case of a sudden perturbation (such as an instantaneous change in the load somewhere in the network) arising in an otherwise quiescent solution, bearing in mind that it takes time to adapt the time step. By limiting the maximum step size to a problem-specific value and by enabling gradual but sufficiently fast decreasing time steps, one may also accommodate these situations.

Three of the adaptive methods considered here (RK23, RK45 and DOP853) are explicit, meaning that every iteration is computed from known information. The adaptivity of these methods is based on so-called embedded error estimates. This means that the methods compare the numerical solution obtained with Runge-Kutta (RK) methods of different order of convergence to infer whether the difference between these numerical solutions is below a certain tolerance level. If the selected norm of the difference between the involved RK methods is below the tolerance, the time step size is increased and if it is above, the time step size is decreased. All methods considered here have the First Step As Last (FSAL) property, which means that the lower-order RK scheme is contained in the higher order one. The FSAL property means that the adaptive scheme is as expensive as its highest-order RK scheme. So for example, one could compute the steps associated with a fifth-order RK scheme and compare it with a fourth-order RK scheme to determine how to change the time step. Hence, one order of convergence is traded to gain adaptive time-stepping.

The remaining two methods (BDF and Radau) are implicit, which means that iterative methods are required each time step. The necessity for iterative solvers each time step means that implicit methods are in general more expensive per time step, but they have much better stability properties with regard to time-step size compared to explicit methods. The implicit adaptive schemes can adapt the time-step size much quicker than the explicit schemes as a result of their better stability. This means that while implicit methods are more expensive per time step, they can save computational costs for cases that overall require significantly fewer steps compared to explicit schemes. The Newton-Raphson method is the default iterative solver in the implicit methods and crucial computational cost is saved by supplying the Jacobian \eqref{eq:Jacobian} of the system in a sparse format to the implicit methods. If one neglects sparsity, one quickly runs into memory issues for merely assembling the Jacobian.

We now discuss the five adaptive methods in more detail.

\paragraph{RK45.} The RK45 method is based on an explicit Runge-Kutta pair of orders 4 and 5 and has the FSAL property. The method was developed in \cite{dormand1980family} and is the default method in many software packages for solving initial value problems. It is the default method in scipy (\cite{virtanen2020scipy}) and is the method behind MATLAB's (\cite{MATLAB}) ode45.
\paragraph{RK23.}
RK23 is based on an explicit Runge-Kutta pair of orders 2 and 3 and has the FSAL property. It is based on the work of \cite{bogacki19893} and is designed to be more efficient than the RK45 method at crude tolerances and in the presence of moderate stiffness. In MATLAB, the function ode23 is based on the RK23 method.
\paragraph{DOP853.} The DOP853 method can be found in \cite{hairer1993solving} and is an eighth order method used for problems where high accuracy is necessary, i.e., for problems where the absolute and relative tolerances are required to be small. 
\paragraph{BDF.} The BDF method is an implicit multistep variable order (one to five) method based on the backward differentiation formula. Its implementation in MATLAB and scipy can be found in \cite{shampine1997matlab}. It is suitable for solving stiff ordinary differential equations and is based on approximating derivatives of the solution. Hence the BDF method is not suitable for systems with discontinuous behaviour.
\paragraph{Radau.} The Radau method is a fifth order implicit RK scheme based on the Radau quadrature. It is the method behind MATLAB's ode23s function. The Radau method is suitable for stiff problems and has a high order of accuracy.

In this section we introduced the mathematical formulation for a DC network built out of distributed generation units that are connected by physical lines. We formulated the control objectives and the control law that guarantees the desired behaviour of the network. By investigating the structure of the affine system of differential equations that describe the DC network, five adaptive numerical methods are identified that will be used in the next section to solve the system \eqref{eq:closedloop} numerically for different network topologies.

\section{DC network simulations}\label{sec:numerics}
In this section, we present simulation results of DC networks based on realistic AC network topologies using the adaptive methods introduced in the previous section. Since DC networks and especially large DC networks have not received as much attention as AC networks, there are only few realistic DC network topologies available in literature. Hence we choose to use the topologies corresponding to realistic AC networks and numerically solve the system \eqref{eq:closedloop} of differential equations based on these topologies.

Realistic AC power network configurations are publicly accessible through the Python package pandapower, see \cite{thurner2018pandapower}. The package provides the topology of the networks and links to algorithms from graph theory through its compatibility with the Python package networkx, see \cite{hagberg2020networkx}. These tools allow the analysis of pandapower networks and in particular a straightforwar extraction of important graph-theoretical matrices, such as the incidence and adjacency matrices, for the pandapower networks. In addition to the network structure, pandapower provides the details and characteristics of the elements in the network, together with algorithms to solve power flow equations in the frequency domain. For our numerical experiment, we turn AC networks into DC networks in the following manner. In the context of AC networks, one considers buses (nodes) and transmission lines (edges), whereas in DC networks context, the nodes are DGUs. In the AC networks available in pandapower, some of the buses are generators, while other buses consume power. Inspired by \cite{de2018power}, we use this AC information to set up the communication network. More precisely, the communication network has the same node set as the physical network, but only generator nodes are communicating with each other. As in the DC network, both generators and consumers are modelled by DGUs, we distinguish between generators and other nodes by setting the parameters for the representing DGUs differently, (for details, see Table \ref{tab:params} in the Appendix). For the communication network, we use a ring structure among the generators in the original AC network we interconnect nodes that are generators in the original AC power network in a nearest neighbour fashion. This implies that the Laplacian matrix $\mathcal{L}^{com}$ of the communication network is the Laplacian matrix of a cycle graph with the number of nodes equal to the number of generators and thereby sparse. The construction of a communication network in this way closely follows that of \cite{de2018power}, where power-consensus algorithms for DC networks are developed. This procedure turns an AC power network into a DC network.

We generate the parameters of the DGUs and the lines of the network uniformly randomly (for details see Table \ref{tab:params} in the Appendix). This choice is based on two reasons. Firstly, in this way, we can represent realistic situations, where similar devices will have similar values but seldom have identical values. Secondly, the uniformly random selection of the parameters is a simple and reproducible way to have single rule that applies to networks of different sizes. In the next section, we explain the computational experiment that we executed for each of the 26 networks described in Table \ref{tab:networks} in the Appendix with each of the 5 adaptive methods.

\subsection{Simulations}
In the following section, we describe the computational experiment that we designed for a fair comparison between the adaptive numerical methods. The same computational experiment is executed for each DC network and each adaptive method as follows. Starting from the initial conditions given in Table \ref{tab:ics} in the Appendix, we simulate each network for a long period of time with each of the five methods. We chose this period to be 5 seconds since the initial conditions are taken random, and by that, with high probability, the network is not in a steady state and the control objectives of current balancing and average voltage regulation are not satisfied in the initial state. After a transient period that is typically around 0.1s for the parameters provided in \cite{trip2018distributed}, the voltage of the network settles to the prescribed values in $\boldsymbol{V}^*$. The currents have a transient of about a second before achieving the current balancing control objective. All networks relax to a steady state that satisfies the control objectives in more or less the same time, with the longest transients being a little longer than a second. Hence a three second simulation would suffice for studying the behaviour after initialising in a nonequilibrium state. However, we introduce two perturbations to study the response of the adaptive time-stepping methods. At time $t=1.5$, we set the load of the first DGU that is not a generator to 20A. This leads to a short period of voltage undershoot. At $t=2.0$ we remove the added load, leading to a brief voltage overshoot after which the network again settles into an equilibrium that satisfies the control objectives. In this way, the adaptive scheme has to adjust its step size twice to deal with the perturbations. This is important for a fair comparison between explicit and implicit adaptive schemes. Before we present this comparison, we first make the following remark. 

Note that it is also possible to initialise in a state that is steady and satisfies the control objectives using the approach provided in \cite{trip2018distributed}. However, this approach relies on the computation of pseudoinverses of matrices which become prohibitively expensive to compute for large networks. Instead, we let the network settle itself into an equilibrium that satisfies the control objectives. 

\begin{figure}
    \centering
    \includegraphics[width=\textwidth]{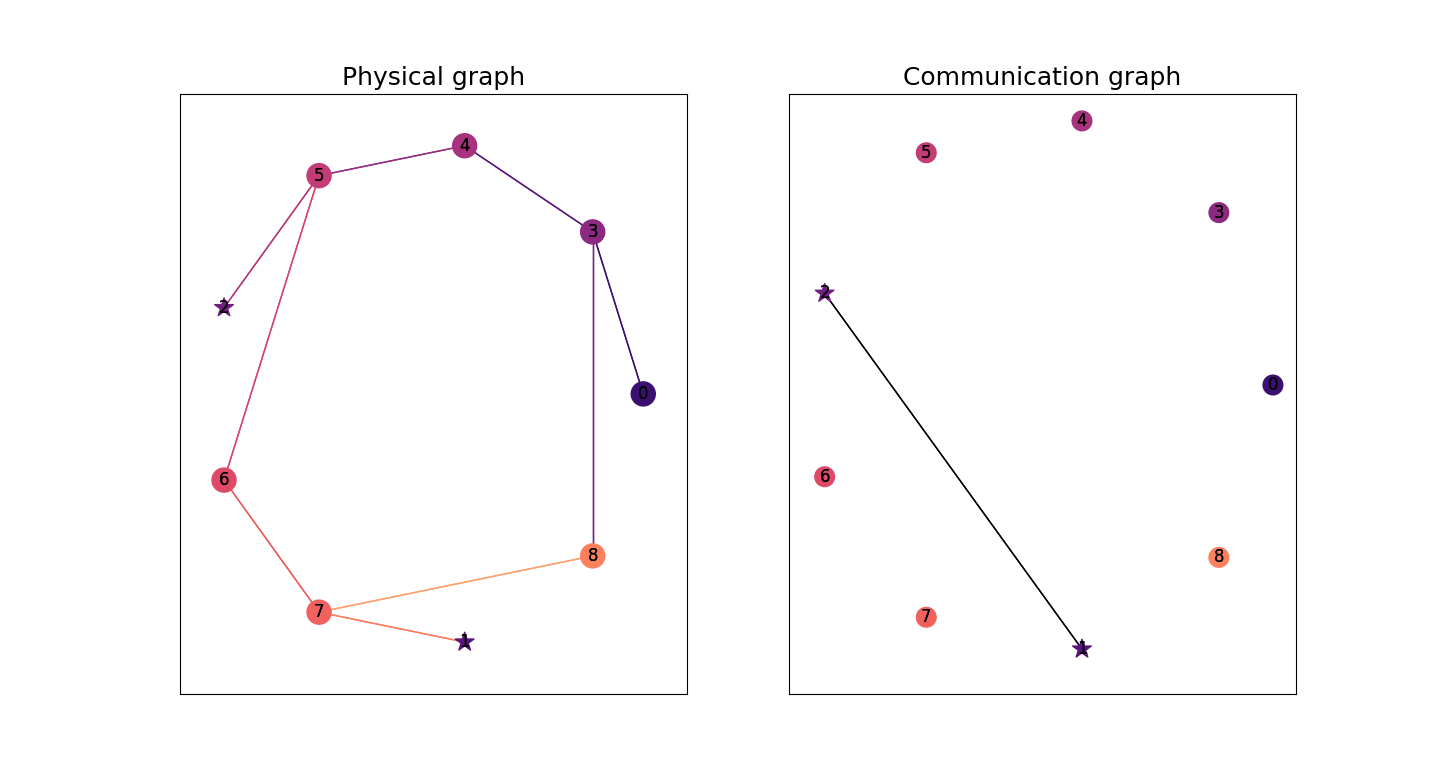}
    \caption{The physical graph and communication graph of the IEEE Case 9 network, plotted in the circular layout. Generator nodes are indicated with a star. The color scale is used to visualise the voltage and currents in Figure \ref{fig:case9response}.}
    \label{fig:case9networks}
\end{figure}

\begin{figure}
    \centering
    \includegraphics[width=\textwidth]{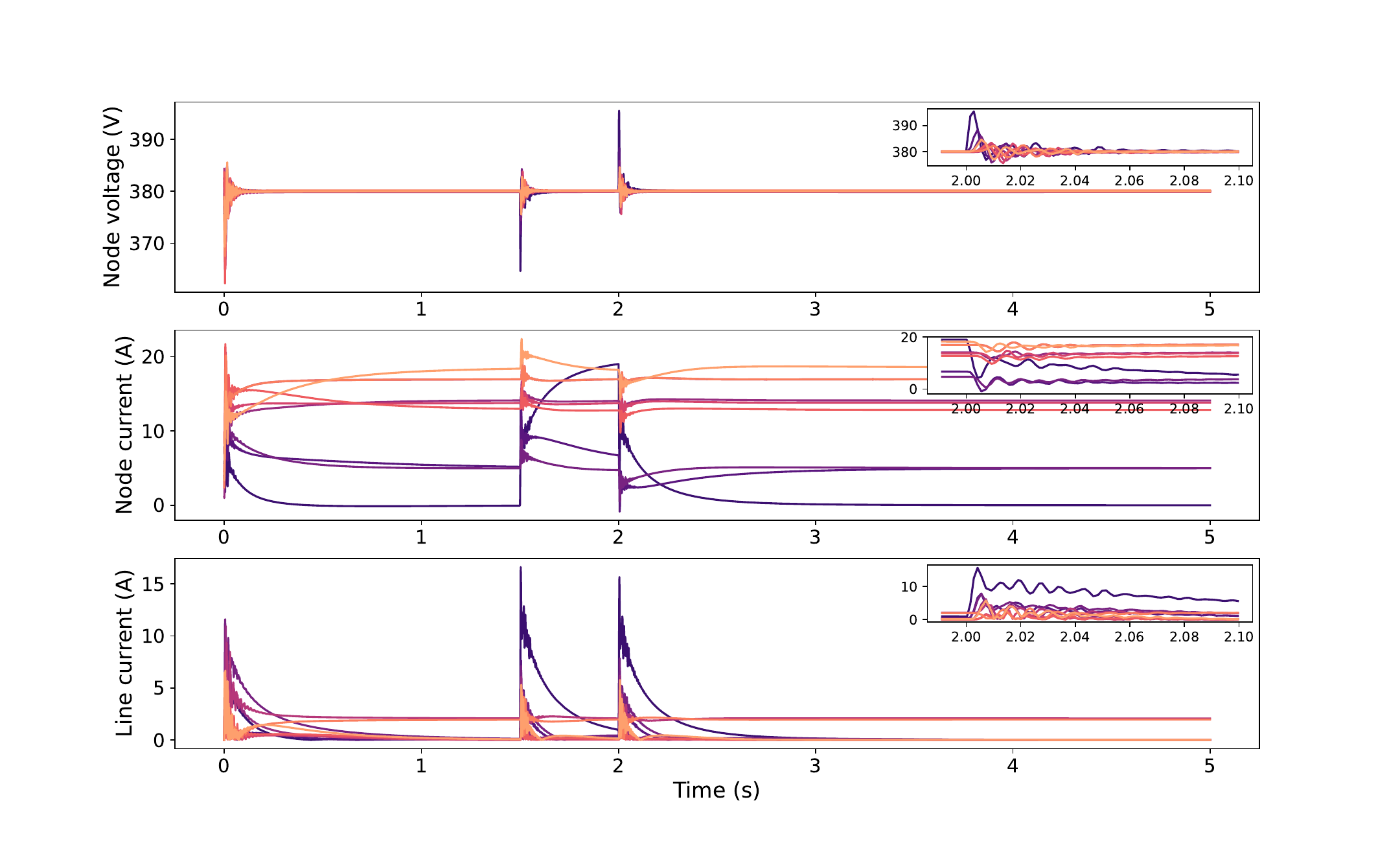}
    \caption{Simulation results of the computational experiment for the IEEE Case 9 network. The colors match with the nodes and lines in Figure \ref{fig:case9networks}. The top plot shows the voltage at the point of common connection for each DGU. The middle plot shows the generated current for each DGU. The bottom plot shows the absolute value of the line currents for each line. The initial state does not satisfy the control objectives, but after a short time, the state converges to an equilibrium that does satisfy the objectives of proportional current sharing and average voltage regulation. At time $t=1.5$, the load is increased in one node, and at time $t=2.0$, the load is removed.}
    \label{fig:case9response}
\end{figure}

Before we present the comparison, we first present an example. We consider the network IEEE Case 9, which consists of 9 nodes, 9 edges, and 2 generators. The graphs of the physical network and the communication network are shown in Figure \ref{fig:case9networks}. In Figure \ref{fig:case9response}, the results of the computational experiment for IEEE Case 9 are given. The nodes and edges in Figure \ref{fig:case9networks} are colored to support their identification in Figure \ref{fig:case9response}. Note in particular that in Figure \ref{fig:case9response} the times for the transients to subside for the node and line current are considerably longer than the time for the transients to subside in the node voltage. This is further motivation for the relatively long simulation time compared to literature.


We performed this computational experiment for all 26 networks listed in Table \ref{tab:networks} in the Appendix and record the number of solver evaluations and the time-dependent time-step size throughout the computation. The number of solver evaluations is a means to track how well a numerical method adapts to the behaviour of the solution. A large number of solver evaluations indicates that the numerical method had difficulty in following the behaviour of the solution, which in general leads to longer simulation times. In the comparison of the methods, this is where the difference between explicit and implicit becomes particularly interesting. Explicit methods are cheaper per time step than implicit methods but do not adapt their step size as fast. The computational cost is expressed here in terms of simulation time, which is a hardware-dependent quantity\footnote{The simulations are performed on a 2020 MacBook Pro with a 2GHz Quad-Core Intel i5 processor and 16GB memory.}. However, the qualitative behaviour of the simulation time per network is a proper indication of the performance of a method. In other words, if one method is twice as expensive as another method on the machine used for the computations, we expect the same behaviour to be the case on other machines.  In the next section, we show the complexity of the numerical methods for the experiment described.

\subsection{Complexity}
\begin{figure}
    \centering
    \includegraphics[width=.8\textwidth]{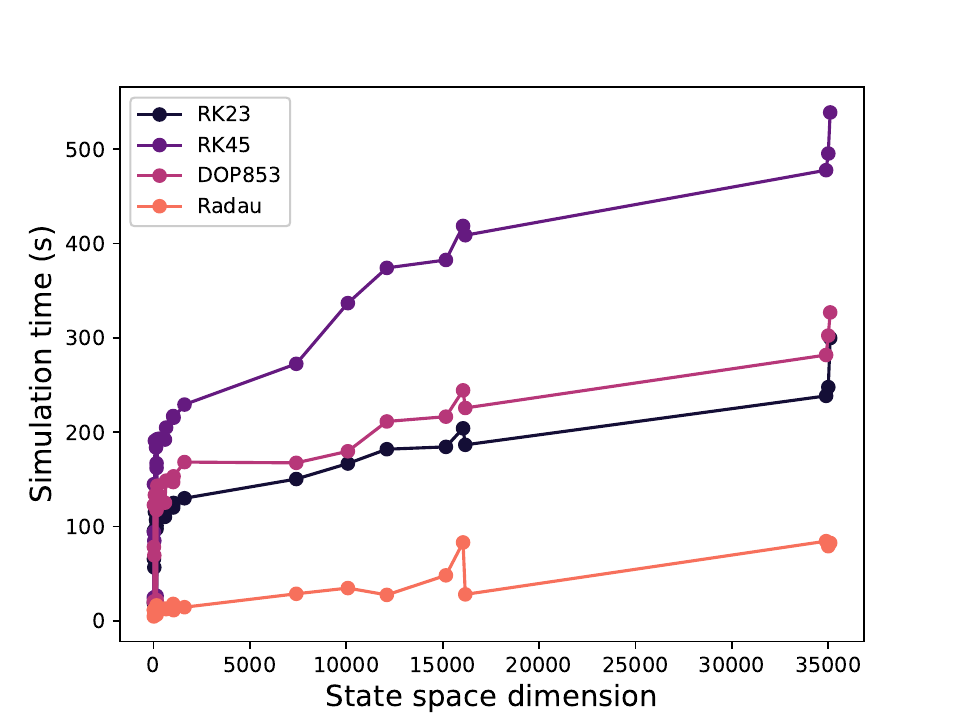}
    \caption{Simulation results of the DC networks based on power network topologies. For the competitive adaptive schemes, the simulation time is shown versus the state space dimension associated with each network.}
    \label{fig:simtime}
\end{figure}

In this section we discuss the results of the computational experiment for all the networks. We focus on the qualitative behaviour of the computational effort to solve the DC network equations \eqref{eq:closedloop} as a function of the state-space dimension of the network. In particular, how the computational cost depends on the particular adaptive numerical method that was used, the size of the network, and the number of generators. Although the simulation time itself is a hardware-dependent quantity, it gives a qualitative indication of how the computational cost grows with the number of differential equations (see Figure \ref{fig:simtime}).

The following observations can be made.
\begin{itemize}
    \item The BDF method is excluded from Figure \ref{fig:simtime} because it is not competitive. This is to be expected since the BDF method cannot handle discontinuous signals, so it struggles with the attachment and detachment of the load in our computational experiment.
    \item The most commonly used method for numerical solving of differential equations is RK45, and its performance is the poorest out of the remaining methods. This is because RK45 was not designed to handle stiff systems of differential equations, which emerge here due to the slow decay of transients versus the fast oscillations of the LC component of the network.
    \item RK23 is specifically suited for stiff systems and is faster than DOP853. However, the high accuracy of the DOP853 explicit Runge-Kutta method gives it a competitive edge over RK23 in our applications which is why one may consider DOP853 the better choice for the simulation of DC networks.
    \item  The Radau method is significantly better in terms of performance than the other methods, by a generous margin of a factor 2-5 compared to the competitors DOP853, RK23 and RK45, respectively (cf. Figure \ref{fig:simtime}). 
\end{itemize}  

In all cases, i.e., all networks and all time integration methods, one recognises a more or less linear scaling between the dimension of the state space and the simulation time. The key to this behaviour is using sparsity throughout. Indeed, the complete embedding of sparsity in the numerical implementation of the system of the equations is the core reason why simulation of such high-dimensional problems with conventional methods is possible in the first place.

The number of generators also has a significant impact on the total simulation time, in particular for the larger networks. In Table \ref{tab:networks}, it can be seen that network Case2869pegase has a significantly larger number of generators than networks of similar size. This explains the increase in simulation time at state space dimension 16058, which corresponds to Case2869pegase. More importantly, the increase in simulation time is the largest for the Radau method. In Table \ref{tab:networks}, it can be seen that network Case9241pegase, which is the largest network in this study, also has proportionally many more generators than the other large networks. We have excluded this result from the simulation time results in Figure \ref{fig:simtime}, but from Table \ref{tab:networks} we can also conclude that the three explicit methods continue their linear behaviour, whereas the Radau method changes its behaviour and becomes the worst of the four in terms of simulation time. This is likely due to the proportionally large number of generators in Case9241pegase.

\section{Outlook and conclusion}\label{sec:outlookandconclusion}
In this work, a numerical tool for DC networks is developed and used to simulate electricity networks varying from small to very large. By exploiting in the implementation the sparsity of the networks, conventional time integration methods are very practical. To evaluate the proposed tool, we used the IEEE benchmark topologies, which all have sparse incidence matrices. Inspired by \cite{de2018power}, the communication network was represented by a ring structure over all generators in the network, which led to a sparse Laplacian matrix for the communication network. By further implementing all parameters and variables in a sparse manner, we are able to use conventional adaptive time integration methods to solve the system of stiff differential equations with discontinuous signals. 

We showed that the computational cost of the methods scales linearly with the size of the network for explicit adaptive time-stepping methods and that out of the explicit methods, RK23 is the fastest. However, at a slightly higher computational cost, the DOP853 method is able to reach much more accurate results. The default method of many software packages is RK45, which produces spurious oscillations when the networks become larger and is the slowest explicit method. The implicit methods do not scale linearly in terms of the computational cost . The BDF method is unsuitable for discontinuous signals and is not able to compete with the other methods in terms of simulation time. The Radau method has the lowest overall computational cost because of its excellent ability to deal with stiff systems. However, for the largest network, the Radau method loses its excellent performance, likely due to its sensitivity to the large number of generators. Hence, we conclude that the DOP853 is the best choice for general DC networks because of its robustness as the networks become larger and that the Radau method is the best overall provided that the networks are not too large and do not have a large number of generators.

In future work, the proposed numerical infrastructure can be used for fault simulation and optimisation of communication infrastructure. For fault simulation in networks of moderate size, the Radau method is best suited. For the optimisation of transient behaviour using communication networks, DOP853 is the better choice, due its robustness to changes in the communication network. In particular, for usingsimulation-based optimisation, the computational cost is an essential factor and this is where the present results provide guidance. Another important research direction includes the design of passive models for the stable interconnection of DC networks with AC networks, as indicated by \cite{sahoo2017control}. 

\section*{Appendix}\label{app:params}
In Table \ref{tab:params} $\mathbb{1}_n$ denotes an $n$-vector of ones and $U([\alpha,\beta],n)$ refers to an $n$-vector of uniformly random numbers with minimum $\alpha\in\mathbb{R}_+$ and maximum $\beta\in\mathbb{R}_+$. The specific values are taken comparable to \cite{trip2018distributed2}.
\begin{table}[H]
    \centering
    \begin{tabular}{l|c|c|c}
       \multirow{2}*{Operation parameters} & $\mathbf{V}^*$  & $380\, \mathbb{1}_n$ & V \\
       & $\mathbf{I}_L$  &  $U([10,20], n-n_{gen})$ & A\\
     \hline \multirow{3}*{Network parameters} & $B$ &  & \\
        & $B^{com}$ & & \\
       & $\mathcal{L}^{com}$ & $B^{com}\Gamma(B^{com})^T$ &\\
       \hline \multirow{2}*{Node parameters} & $L_t$ & $U([1.5,3.5],n)$ & mH\\
       & $C_t$ & $U([1.5,2.5],n)$ & mF\\
       \hline \multirow{2}*{Line parameters} & $R$ & $U([40,100],m)$ & $\Omega$\\
       & $L$ & $U([1.5,2.5],m)$ & mH\\
       \hline \multirow{5}*{Controller parameters}& $T_\theta$ & $\mathbb{1}_n$ &\\
       & $T_\phi$ & $10^{-2}\,\mathbb{1}_n$ & \\
       & $K$ & $\frac{1}{2}\,\mathbb{1}_n$ & \\
       & $W$ & $\mathbb{1}_n$ & \\
       & $\Gamma$ & $10^2\, \mathbb{1}_{m^{com}}$ & \\
    \end{tabular}
    \caption{DC network parameters}
    \label{tab:params}
\end{table}
The initial conditions are taken uniformly random, with the initial generated current between 0A and 10A and the initial voltage between 375V and 385V, as in Table \ref{tab:ics}. The initial line current and controller variables are set to the zero vector.
\begin{table}[H]
    \centering
    \begin{tabular}{c|c}
        $\boldsymbol{I}_0$ & $U([0,10],n)$ \\
        $\boldsymbol{V}_0$ & $U([370,390],n)$\\
        $\boldsymbol{f}_0$ & $\mathbb{0}_m$\\
        $\boldsymbol{\theta}_0$ & $\mathbb{0}_n$\\
        $\boldsymbol{\phi}_0$ & $\mathbb{0}_n$
    \end{tabular}
    \caption{Initial conditions}
    \label{tab:ics}
\end{table}
We select a network from Table \ref{tab:networks}.
Except for $\mathbf{V}^*, \mathbf{I}_L$ and $\mathcal{L}^{com}$, all other parameters in the table are diagonal matrices and can therefore be specified by a single vector. This sparsity structure is crucial when dealing with large networks to avoid memory issues when assembling the Jacobian associated with \eqref{eq:closedloop}. 

\begin{table}[H]
    \centering
    \begin{tabular}{l|cc|c|c|c}
       Network  & $\#$ of nodes $(n)$ & $\#$ of edges $(m)$ & \begin{tabular}{@{}c@{}} average $\#$ edges\\
       per node\end{tabular} & $\#$ of generators & \begin{tabular}{@{}c@{}} dimension of \\ state space $(4n+m)$\end{tabular}  \\
       \hline Case 4gs  & 4 & 4 & 2.00 & 1 & 20\\
       Case 5 & 5 & 6 & 2.40 & 3 & 26\\
       Case 6ww & 6 & 11 & 3.67 & 2 & 35\\
       Case 9 & 9 & 9 & 2.00 & 2 & 45\\
       Case 14 & 14 & 20 & 2.86 & 4 & 76\\
       Case 24 ieee rts & 24 & 38 & 3.17 & 10 & 134\\
       Case 30 & 30 & 41 & 2.73 & 5 & 161\\
       Case IEEE 30 & 30 & 41 & 2.73 & 5 & 161\\
       Case 33bw & 33 & 32 & 1.94 & 0 & 164\\
       Case 39 & 39 & 46 & 2.36 & 9 & 202\\
       Case 57 & 57 & 80 & 2.81 & 6 & 308\\
       Case 89pegase & 89 & 210 & 4.72 & 11 & 596\\
       Case 118 & 118 & 186 & 3.15 & 53 & 658\\
       Case 145 & 145 & 453 & 6.25 & 49 & 1025\\
       Case Illinois 200 & 200 & 245 & 2.45 & 37 & 1045\\
       Case 300 & 300 & 411 & 2.74 & 68 & 1611\\
       Case 1354pegase & 1354 & 1991 & 2.94 & 259 & 7407\\
       Case 1888rte & 1888 & 2531 & 2.68 & 271 & 10083\\
       GB network & 2224 & 3207 & 2.88 & 393 & 12103\\
       Case 2848rte & 2848 & 3776 & 2.65 & 369 & 15168 \\
       Case 2869pegase & 2869 & 4582 & 3.19 & 509 & 16058\\
       Case 3120sp & 3120 & 3693 & 2.37 & 247 & 16173\\
       Case 6470rte & 6470 & 9005 & 2.78 & 452 & 34885 \\
       Case 6495rte & 6495 & 9019 & 2.78 & 487 & 34999\\
       Case 6515rte & 6515 & 9037 & 2.77 & 492 & 35097\\
       Case 9241pegase & 9241 & 16049 & 3.47 & 1444 & 53013 
    \end{tabular}
    \caption{Power system networks available in pandapower \cite{thurner2018pandapower} and their number of nodes, edges, generators, and state space dimension when used in the closed loop system of differential equations for the DC network.}
    \label{tab:networks}
\end{table}

\bibliographystyle{plainnat}
\bibliography{biblio}

\end{document}